\documentclass[12pt]{iopart}
\usepackage{graphics}
\usepackage{url}
\usepackage{color}

\begin{document}

\title{Bayesian inference of  the incompressibility, skewness and kurtosis of nuclear matter from empirical pressures in relativistic heavy-ion collisions}

\author{Wen-Jie Xie$^{1}$ and Bao-An Li$^{2}$\footnote{Corresponding author: Bao-An.Li@Tamuc.edu} }
\address {$^{1}$Department of Physics, Yuncheng University, Yuncheng 044000, China}
\address {$^{2}$Department of Physics and Astronomy, Texas A\&M University-Commerce, TX 75429-3011, USA}

\date{\today}
\setcounter{page}{1}
\begin{abstract}
Within the Bayesian statistical framework we infer the incompressibility $K_0$, skewness $J_0$ and kurtosis $Z_0$ parameters of symmetric nuclear matter (SNM) at its saturation density $\rho_0$ using the constraining bands on the pressure in cold SNM in the density range of 1.3$\rho_0$ to 4.5$\rho_0$ from transport model analyses of kaon production and nuclear collective flow in relativistic heavy-ion collisions.
As the default option assuming the $K_0$, $J_0$ and $Z_0$ have Gaussian prior probability distribution functions (PDFs) with the means and variances of $235\pm 30$, $-200\pm 200$ and $-146\pm 1728$ MeV,
their posterior most probable values are narrowed down to 192$^{+12}_{-16}$ MeV, -180$^{+100}_{-110}$ MeV and 200$^{+250}_{-250}$ at 68\% confidence level, respectively. The results are largely independent of the prior PDFs of $J_0$ and $Z_0$ used. However, if one adopts the strong belief that the incompressibility $K_0$ has a uniform prior PDF within its absolute boundary of 220-260 MeV as one can find easily in the literature, the posterior most probable values of $K_0$, $J_0$ and $Z_0$ shift to $K_0=220^{+6}_{-0}$ MeV, $J_0=-390^{+60}_{-70}$ MeV and $Z_0=600^{+200}_{-200}$ MeV, respectively. While the posterior PDFs of the SNM EOS parameters depend somewhat on the prior PDF of $K_0$ used, the results from using different prior PDFs are qualitatively consistent. The uncertainties of all three parameters are significantly reduced especially for the $J_0$ and $Z_0$ parameters compared to their current values.

\end{abstract}

\submitto{\jpg}
\maketitle
\section{Introduction}
Thanks to the great efforts over the last 4 decades by many people, see, e.g., Ref. \cite{Bla1} for an earlier review, the incompressibility $K_0=9\rho_0^2[d^2E_0(\rho)/d \rho^2]_{\rho_0}$ measuring the stiffness of symmetric nuclear matter (SNM) Equation of State (EOS) $E_0(\rho)$ around its saturation density $\rho_0$ has been relatively well determined to be about 240$\pm$20 MeV \cite{Young,shlomo06,Jorge10,Garg18} or 230$\pm$40 MeV \cite{Khan1,Khan2,Khan3} while there is a report of somewhat higher values in the range of $250\leq K_0\leq 315$ MeV \cite{Stone} mostly based on systematic studies of the available Giant Monopole Resonance (GMR) data of some heavy nuclei. It has been pointed out by several groups that the main sources of the remaining uncertainties and model dependences in pinning down the $K_0$ further is its correlations with the uncertain high-order density dependence of both the  $E_0(\rho)$ and nuclear symmetry energy $E_{\rm{sym}}(\rho)$ \cite{Khan1,Khan2,Khan3,Colo08,Colo14,MM1,Margueron19}.

Unfortunately, the stiffness of SNM EOS at supra-saturation densities characterized by the skewness parameter $J_0=27\rho_0^3[d^3E_0(\rho)/d \rho^3]_{\rho_0}$ and the kurtosis parameter $Z_0=81\rho_0^4[d^4E_0(\rho)/d \rho^4]_{\rho_0}$ is hardly known. As pointed out already by Margueron et al. \cite{MM1}, there were only few estimations of the poorly known $J_0$ from analyzing experimental data. Moreover, most of the gross properties and GMR of finite nuclei are only sensitive to the EOS near the so-called crossing-density of about 0.10 fm$^{-3}$ \cite{Khan1,Khan2,Khan3,Cen09,Dan09,Che11a,Jorge11,Zha15}.
In fact, even its sign is not determined firmly. For the latest and most comprehensive review of model predictions for $J_0$ in the range of -369 MeV to 1488 MeV, we refer the reader to Ref. \cite{MM1}. In particular, negative values of $J_0$ were suggested by some non-relativistic Skyrme and/or Gogny Hartree-Fock calculations \cite{MM1,Dutra12,sellahewa14,Far97,Lwchen11}, relativistic mean-field models \cite{Cai14} as well as several analyses of some neutron-star observations \cite{Steiner10,Zhang18apj,Zhang19apj,Xie19,Xie20,Zhou19a,Zhou19b}. For example, considering the constraints on the pressure of SNM imposed by both the flow data in heavy-ion collisions \cite{Dan02} and the mass of PSR J0348+0432 \cite{Antoniadis13}, a range of -494 MeV $\leq J_0 \leq$ -10 MeV was inferred within a nonlinear relativistic mean field model \cite{Cai14}. While $J_0=-190_{-40}^{+40}$ MeV at 68\% confidence level was found in our recent Bayesian analysis \cite{Xie19,Xie20} of neutron star radii from X-ray observations and the tidal deformability of GW170817 under the constraints of causality and reproducing the maximum mass of neutron stars at least as high as $M$=2.17$_{-0.10}^{+0.11}$ M$_{\odot}$ as indicated by the first report \cite{M217} of the mass of PSR J0740+6620 \cite{M214}. On the other hand, positive values of $J_0$ were predicted by some other relativistic mean field models \cite{MM1,Dutra14}. For example, within a relativistic density functional theory constrained by both terrestrial experiments and astrophysical observations as well as predictions of chiral effective field theories at low densities, very large values of $J_0$ in the range of 300 to 800 MeV were predicted \cite{Jiajieli19}, going beyond the already large range of approximately -800 MeV $\leq J_0 \leq$ 400 MeV previously known from surveying earlier analyses of terrestrial experiments and astrophysical observations as well as predictions of over 500 nuclear energy density functionals \cite{Tews17,Zhang17}.

It is very interesting to note that more efforts are constantly being made by the nuclear physics community to both understand why the $J_0$ parameter is so poorly known and how to better determine it. For example, a recent study in the framework of the Landau-Migdal theory shows that three-particle correlations play a crucial role in determining the value of $J_0$ \cite{Bentz19}. This is consistent with earlier findings within Skyrme/Gogny Hartree-Fock calculations that the $t_3$ term charactering effectively the density dependence of many-body interactions/correlations is important but poorly understood for determining the $K_0$ and $J_0$ parameters as well as their correlations \cite{Khan1,Khan2,Khan3,MM1}. Moreover, the latest and state-of-the-art Quantum Monte Carlo calculations using local interactions derived from chiral effective field theories up to the next-to-next-to-leading order found a value of $252 \pm 390 \leq J_0\leq 1110\pm 1491$ MeV depending on the parametrization of the three-body force used within the statistical Monte Carlo errors and the uncertainties coming from the truncation of the chiral expansion \cite{Diego}.

\begin{table}[htbp]
\centering
\caption{\label{tab:priornew} The Gaussian prior parameters and boundaries for the EOS parameters $K_0$, $J_0$ and $Z_0$ in MeV, where the Av and $\sigma$ denote the averages and variances of the Gaussian distributions, respectively.}
 \begin{tabular}{lccc}
  \hline\hline
  & $K_0$ &$J_0$ &$Z_0$ \\
  \hline\hline\\
 Av  &235 &-200 &-146  \\
 $\sigma$ &30 &200 &1728 \\
 Min &145 &-800 &-5330\\
 Max &325 &400 &5038\\
 \hline\hline
 \end{tabular}
\end{table}

The kurtosis parameter $Z_0$ (denoted by $Z_{\mathrm{sat}}$ elsewhere, e.g., see Refs.\cite{MM1,Antie19}) plays an important role at densities higher than about 3$\rho_0$. It is even more poorly known than $J_0$ as one expects.
It is not always considered in parameterizing the SNM EOS not only because it is very poorly known but also because it starts playing significant roles in density regions where a hadron-quark phase transition is expected.
Since we are going to use the pressure bands extracted from heavy-ion collisions using transport models assuming no such phase transition below $4.5\rho_0$, it is more appropriate to consider the $Z_0$ term in the present work.
As discussed in Refs. \cite{MM1,Antie19}, the uncertainty range for $Z_0$ is very wide. For example, a range of 901 MeV $\leq Z_0 \leq$ 1537 MeV was predicted by the chiral effective field theory \cite{Drischler16} while the empirical range provided by the Skyrme-type interactions is -903 MeV $\leq Z_0 \leq$ 2128 MeV \cite{MM1}. Moreover, a completely negative $Z_0$ range of -4478.35 MeV $\leq Z_0 \leq$ -353.91 MeV was suggested by an empirical local density functional model \cite{MM1}. At the same time, very large positive $Z_0$ values of 2014 MeV $\leq Z_0 \leq$ 9997 MeV and 4581 MeV $\leq Z_0 \leq$ 6703 MeV are used in the relativistic mean field and the relativistic Hartree-Fock approaches \cite{MM1}, respectively. In this study, we adopt a Gaussian prior PDF for $Z_0$ with a mean value of -146 MeV and a variance of 1728 MeV from Ref. \cite{Antie19} in the range of -5330 MeV to +5038 MeV as indicated in Table \ref{tab:priornew}.

Given the current situation mentioned above, much more investigations on the physics associated with the stiffness of dense SNM are obviously necessary. In this work, within the Bayesian statistical framework using well established constraining bands on the cold SNM pressure from transport model analyses of relativistic heavy-ion collisions \cite{Dan02,Fuchs,lynch09}, we infer the posterior PDFs of $K_0$, $J_0$ and $Z_0$ as well as their correlations with their priors as listed in Table \ref{tab:priornew} as the default option. Their most probable posterior values are found to be $K_0=192^{+12}_{-16}$ MeV, $J_0=-180^{+100}_{-110}$ MeV and $Z_0=200^{+250}_{-250}$ MeV at 68\% confidence level, respectively, representing significant refinements to their current values and may serve as a bench mark for future studies on the EOS of super-dense nuclear matter. We also investigate how these results may be altered by using other currently acceptable prior PDFs reflecting the diverse opinions in the field.

The rest of the paper is organized as follows. In Section \ref{app}, we first discuss briefly how the empirical pressures of SNM are extracted from transport model analyses of kaon production and nuclear collective flow in heavy-ion collisions at intermediate/relativistic energies as well as the nature of their error bands. We then discuss how we parameterize the EOS of SNM and give the corresponding pressure as a function of density. We also give some technical details on how we perform the Bayesian inference of the EOS parameters using the empirical pressures in SNM. In Section \ref{Res}, we present and discuss our results. Finally, we summarize our main findings and give an outlook.
\begin{figure}[htb]
\begin{center}
\resizebox{0.85\textwidth}{!}{
  \includegraphics{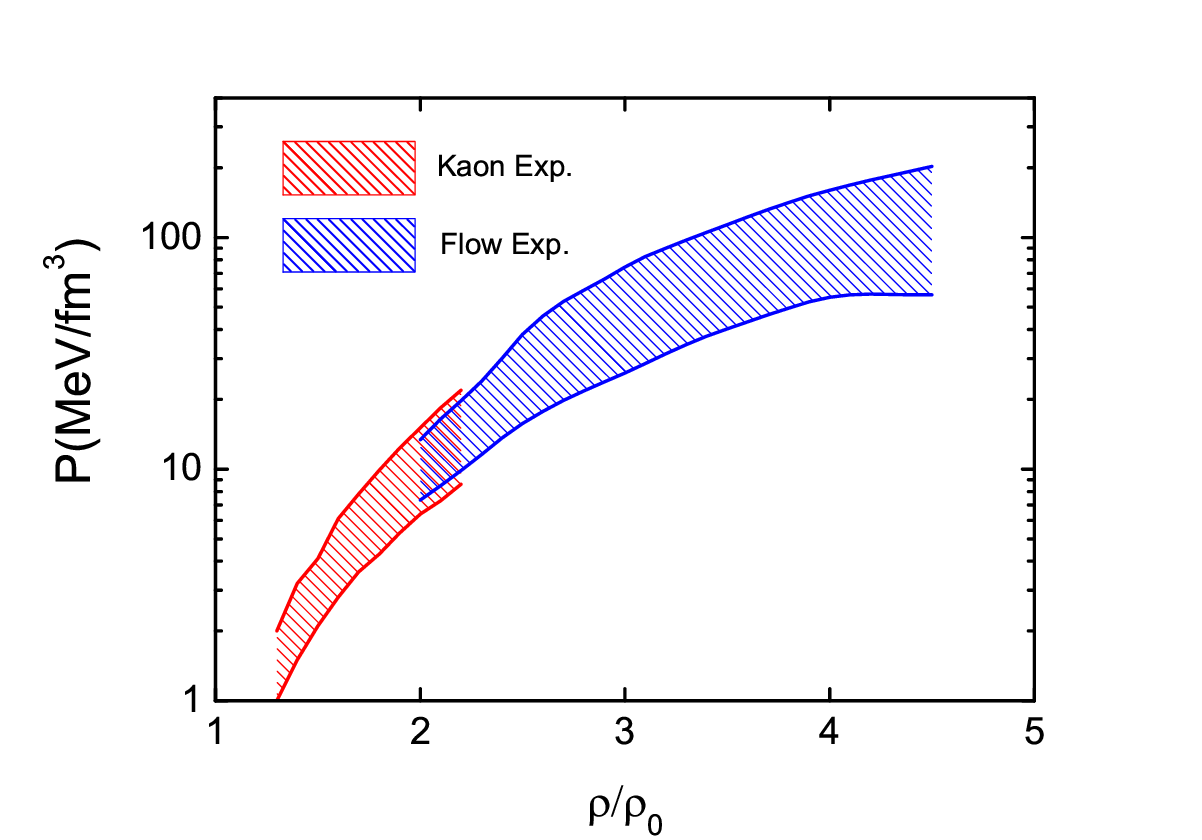}
  }
  \caption{(color online) Constraining bands on the pressure in symmetric nuclear matter as a function of reduced density from analyzing kaon production and nuclear collective flow in energetic heavy-ion collisions. The data are taken from Refs. \cite{Dan02,Fuchs,lynch09}.}\label{pressure}
\end{center}
\end{figure}

\begin{table}[htbp]
\centering
\caption{\label{tab:kaondata} Empirical pressure (MeV/fm$^{3}$) of symmetric nuclear matter with 1$\sigma$ error bar from transport model analyses of kaon production in heavy-ion collisions \cite{Fuchs,lynch09}.}
 \begin{tabular}{lccc}
  \hline\hline
  $\rho/\rho_0$ &Pressure &$\rho/\rho_0$ &Pressure \\
  \hline\hline\\
 1.3 &1.50$\pm$0.33 &1.8 &7.10$\pm$1.87  \\
 1.4 &2.35$\pm$0.57 &1.9 &8.80$\pm$2.33 \\
 1.5 &3.10$\pm$0.67  &2.0 &10.75$\pm$2.90\\
 1.6 &4.45$\pm$1.10  &2.1 &12.85$\pm$3.70 \\
 1.7 &5.70$\pm$1.40  &2.2 &15.25$\pm$4.43\\
 \hline\hline
 \end{tabular}
\end{table}

\begin{table}[htbp]
\centering
\caption{\label{tab:flowdata} Empirical pressure (MeV/fm$^{3}$) of symmetric nuclear matter with 1$\sigma$ error bar from transport model analyses of collective flow in heavy-ion collisions \cite{Dan02}.}
 \begin{tabular}{lccc}
  \hline\hline
  $\rho/\rho_0$ &Pressure &$\rho/\rho_0$&Pressure\\
  \hline\hline\\
 2.0 &10.40$\pm$2.00 & 3.3 &65.90$\pm$20.93  \\
 2.1 &12.50$\pm$2.67 &3.4 &71.25$\pm$22.57 \\
 2.2 &14.85$\pm$3.30  &3.5 &76.95$\pm$24.43\\
 2.3 &17.75$\pm$4.17 &3.6 &82.95$\pm$26.50  \\
 2.4 &21.85$\pm$5.50 &3.7 &89.25$\pm$28.63 \\
 2.5 &26.90$\pm$7.47 &3.8 &95.60$\pm$30.73\\
 2.6 &31.85$\pm$9.37 &3.9 &101.85$\pm$32.83\\
 2.7 &36.30$\pm$11.00  &4.0 &107.60$\pm$35.00 \\
 2.8 &40.45$\pm$12.43 &4.1 &112.55$\pm$37.37\\
 2.9 &45.00$\pm$14.13 &4.2 &116.85$\pm$39.97 \\
 3.0 &50.30$\pm$16.20  &4.3 &120.90$\pm$42.73 \\
 3.1 &55.65$\pm$18.03 & 4.4 &125.05$\pm$45.70\\
 3.2 &60.75$\pm$19.50  &4.5 &129.55$\pm$48.70\\
 \hline\hline
 \end{tabular}
\end{table}

\section{Approach}\label{app}
In this section, we provide some details of our approach.

\subsection{The empirical SNM pressure and its uncertainty from hadronic transport model analyses of relativisitc heavy-ion collisions}
Shown in Fig. \ref{pressure} are the empirical pressures in cold SNM from hadronic transport model analyses of relativistic heavy-ion collisions \cite{Fuchs,lynch09,Dan02}. The results in the density range of 1.3$\rho_0$ to 2.2$\rho_0$ are from studying kaon production \cite{kaos1,kaos2} and  those from 2.0$\rho_0$ to 4.5$\rho_0$ are from studying nuclear collective flow \cite{EOS,E895-1,E895-2,E877}. The means and $1\sigma$ error bars of the two pressure bands are listed in Tables \ref{tab:kaondata} and \ref{tab:flowdata}, respectively.

It is necessary to discuss briefly how the constraining bands on the SNM pressure were obtained. Essentially, they were synthesized from systematic transport model analyses of kaon multiplicities and nuclear collective flows in heavy-ion collisions at intermediate and/or relativistic energies. The upper and lower boundaries in different density regions were set by employing different EOSs with and/or without the momentum dependence of single-nucleon mean-field potentials sometimes within different transport codes \cite{Dan02,Fuchs,lynch09}. The underlying values of $K_0$ used in the original data analyses range from about 170 MeV to 380 MeV depending on if/what kinds of the momentum dependent single-nucleon potentials were used, and also if/what kinds of in-medium nucleon-nucleon cross sections were used. While the underlying $J_0$ and $Z_0$ values of these models were generally not given/known.
The error bands contain combined uncertainties of all theoretical model ingredients and data analyses as well as the experiments themselves. 

The dynamics and observables of heavy-ion collisions are determined by the nuclear effective interactions and correlations in nuclear matter at finite temperature.  
Theoretically, the zero-temperature pressure of SNM is uniquely determined by the $E_0(\rho)$. The corresponding nuclear mean-field potential determined by the same interaction as the $E_0(\rho)$ in cold nuclear matter is a direct input in transport model simulations of heavy-ion collisions. Thus, comparing transport model simulations with experimental observations enabled the extraction of SNM pressure at zero temperature over a large density range, assuming the kinetic part of the nuclear pressure is well understood. Within the Boltzmann transport theory, see, e.g., Ref. \cite{Huang}, it is well known that there is an intrinsic degeneracy between the single-nucleon mean-field potential and the in-medium nuclear cross sections in governing the time evolution of nucleon phase space distribution function. Consequently, different combinations of nuclear mean-field potentials related to the $E_0(\rho)$ and the in-medium nuclear cross sections related to the kinetic pressure built during heavy-ion collisions may reproduce the same observables in heavy-ion collisions \cite{Dan02,Fuchs,lynch09,Bert,Sto86,Cas90,Pawel91,Li98,Bass,ditoro,Li2008,Ono19,Xu19,Maria}. Thus, an accurate extraction of the cold EOS from heavy-ion reaction observables  generally requires reliable knowledge about the in-medium nuclear cross sections. Currently, there are still some uncertainties about the latter. This uncertainty is partially responsible for the still relatively large band width reflecting both experimental and theoretical uncertainties of the pressure in cold SNM shown in Fig. \ref{pressure}. Fortunately, as we shall discuss, the Bayesian statistical analysis provides a natural framework to quantify uncertainties of the model parameters consistent with the uncertainties of the data used. Namely, the credible intervals of the PDFs of the EOS parameters naturally reflect the uncertainties of the pressure bands we used.

\subsection{The parameterization of cold SNM EOS}
To infer the PDFs of $K_0$, $J_0$ and $Z_0$ using the empirical pressures discussed above independent of any particular nuclear theory, we adopt the parameterization for $E_0(\rho)$ as
\begin{equation}\label{E0}
  E_{0}(\rho)=E_0(\rho_0)+\frac{K_0}{2}(\frac{\rho-\rho_0}{3\rho_0})^2+\frac{J_0}{6}(\frac{\rho-\rho_0}{3\rho_0})^3+\frac{Z_{0}}{24}(\frac{\rho-\rho_0}{3\rho_0})^4
\end{equation}
with $E_0(\rho_0)$=-15.9 MeV. It has been widely used in the literature in studying properties of nuclei, neutron stars and heavy-ion collisions, see, e.g., Refs. \cite{MM1,Zhang18apj,Xie19,Jiajieli19,Lat16,Oertel17,Lattimer14,NBZ19a,sagert12,sotani14,Baillot19}.
The corresponding pressure in cold SNM is then
 \begin{equation}\label{pressure0}
  P(\rho)=\rho^2\frac{dE_0(\rho)}{d\rho}=\frac{\rho^2}{\rho-\rho_0}[K_0(\frac{\rho-\rho_0}{3\rho_0})^{2}+\frac{J_0}{2}(\frac{\rho-\rho_0}{3\rho_0})^{3}+\frac{Z_{0}}{6}(\frac{\rho-\rho_0}{3\rho_0})^4].
\end{equation}
Normally, one performs Taylor expansions of given energy density functionals $e(\rho)$ based on some nuclear many-body theories. The third-order derivative of $e_(\rho)$ at $\rho_0$, i.e., $27\rho_0^3[d^3e(\rho)/d \rho^3]_{\rho_0}$, is defined as the skewness of SNM EOS and the fourth-order derivative of $e(\rho)$ at $\rho_0$, i.e., $81\rho_0^4[d^4e(\rho)/d \rho^4]_{\rho_0}$, is defined as the kurtosis of SNM EOS. It is necessary to take the value of the derivative at $\rho_0$ so that contributions from high-order terms in $(\rho-\rho_0)/3\rho_0$ in the Taylor expansion of $e(\rho)$ vanish. As already discussed in detail in Refs. \cite{Zhang18apj,Xie19,NBZ19a}, by design the parameterization of Eq.(\ref{E0}) has the form of a Taylor expansion near $\rho_0$ up to the fourth-order term. While the parameterization itself can be considered as a phenomenological energy density functional, we use it purely as a parameterization in our Bayesian analysis. We emphasize that in Taylor expansions, one first needs a known function. While here we are simply inferring/calibrating the coefficients of a parameterization from/using the empirical pressures from heavy-ion reactions. In this work, the Eq. (\ref{E0}) is not a Taylor expansion of any known function.  But it does asymptotically approach a Taylor expansion of some unknown energy functional in the limit of $\rho\rightarrow \rho_0$. Therefore, one can still use the traditional terminologies, e.g., the incompressibility, skewness and  kurtosis, to describe the $K_0$, $J_0$ and $Z_0$ parameters. Moreover, we can use existing predictions for $K_0$, $J_0$ and $Z_0$ in setting their prior ranges in the Bayesian analyses.

It is also necessary to emphasize that both the parameterization of the SNM EOS $E_0(\rho)$ and the empirical pressures extracted from relativistic heavy-ion collisions using hadronic transport models are valid only under the assumption that there is no hadron-quark phase transition below $4.5\rho_0$. Thus, the $J_0$ ($Z_0$) should be considered as an effective skewness (kurtosis) of nucleonic matter under this assumption. Given the limited amount of empirical pressure data available and its nature, it is unnecessary to introduce higher order terms beyond $Z_0$ in the parameterization of Eq. (\ref{E0}) in the present work.

\subsection{The Bayesian inference of SNM EOS parameters}
To calculate the posterior PDFs of $K_0$, $J_0$ and $Z_0$ as well as their correlation functions within the standard Bayesian approach, we use the Metropolis-Hastings algorithm \cite{Metropolis53,Hastings70} in our Markov-Chain Monte Carlo (MCMC) sampling. The posterior probability $P({\cal M}(K_0,J_0,Z_0)|D)$ that a realization ${\cal M}(K_0,J_0,Z_0)$ of our parametric SNM EOS describes correctly the empirical pressures from heavy-ion reactions denoted by $D$ can be formulated as
\begin{equation}\label{Bay1}
P({\cal M}(K_0,J_0,Z_{0})|D) = C P(D|{\cal M}(K_0,J_0,Z_{0})) P({\cal M}(K_0,J_0,Z_{0})),
\end{equation}
where $C$ is a normalization constant and $P({\cal M}(K_0,J_0,Z_0))$ stands for the prior probability distribution function of the model parameters $K_0$, $J_0$ and $Z_0$.
In the present work, as the default option we adopt the Gaussian form for the prior distributions of each parameter and the final prior is a product of the Gaussian distributions of the three parameters, i.e.,
 \begin{equation}\label{PriorDis}
  P({\cal M}(K_0,J_0,Z_0))=\prod_{i=1}^{3}\frac{1}{\sqrt{2\pi}\sigma_{i}}\exp[-\frac{(P_i-\mathrm{Av}_i)^{2}}{2\sigma_i^{2}}],
\end{equation}
where $\mathrm{Av}_i$ and $\sigma_i$ are the averages and variances of the three parameters, respectively, and are given in Table \ref{tab:priornew}. $P_i$ with $i=1,2,3$ denotes $K_0$, $J_0$ and $Z_0$, respectively, and is generated randomly between their minimum and maximum values given in Table \ref{tab:priornew} according to
\begin{equation}\label{sample}
P_i=P_{\mathrm{min},i}+(P_{\mathrm{max},i}-P_{\mathrm{min},i})x,
\end{equation}
where $P_{\mathrm{min},i}$ and $P_{\mathrm{max},i}$ respectively represent the minimum and maximum values of $P_i$, and $x$ is a random number between 0 and 1. $P(D|{\cal M}(K_0,J_0,Z_0))$ is the likelihood to reproduce the empirical pressure $D$ given the model ${\cal M}(K_0,J_0,Z_0)$. It can be expressed as
\begin{equation}\label{Likelihood}
  P[D|{\cal M}(K_0,J_0,Z_0)]=\prod_{j=1}^{N}\frac{1}{\sqrt{2\pi}\sigma_{\mathrm{D},j}}\exp[-\frac{(P_{\mathrm{th},j}-P_{\mathrm{D},j})^{2}}{2\sigma_{\mathrm{D},j}^{2}}],
\end{equation}
where $N$ is the number of data points used. In digitizing the pressures shown in Fig \ref{pressure} from both the kaon and flow data, we use 0.1 as the bin size for the reduced density $\rho/\rho_0$. We have thus $N$=26 (10) for the pressure from the flow (kaon) data set since the relevant density ranges from 2.0$\rho_0$ to 4.5$\rho_0$ (1.3$\rho_0$ to 2.2$\rho_0$) as indicated in Table \ref{tab:flowdata} (Table \ref{tab:kaondata}). When combining the two data sets (named the combined data), we take the points from the flow data in their overlapping region so that the two data sets remain independent and cover two different density regions, which implies that $N$=33 for the combined data set.

The $\sigma_{\mathrm{D},j}$ represents the 1$\sigma$ error bar of the $j$th data point. To avoid confusion, we stress that the word ``data" here refers to the empirical pressure not the kaon multiplicity and/or strength of collective flow directly measured in the experiments. As we emphasized earlier, the error bands shown in Fig. \ref{pressure} include uncertainties in both the experimental data and theoretical modeling. While these empirical pressures still have large uncertainties, they represent the state-of-the-art of the field. In fact, they have been used widely in the literature over the last two decades to test nuclear many-body theory predictions and calibrate various model parameters. In this work, we use the width of the constraining pressure bands as the 3$\sigma$ error bar (99.7\% confidence interval) since the upper and lower limits were given approximately as the absolute boundaries based on the transport model analyses of the heavy-ion reaction experiments \cite{Dan02}. In the following Bayesian analysis, we refer these constraining bands on the SNM pressures as the empirical pressures since they are not directly measurable experimentally. Listed in Tables \ref{tab:kaondata} and \ref{tab:flowdata} are the mean values and $1\sigma$ error bars of the empirical pressures as functions of the reduced density. We notice that the mean values increase smoothly with density, unlike real experimental data that would fluctuate in accordance with the experimental error bars. Since the band width is relatively large and the mean increases smoothly, the digitization with a bin size of $0.1 \rho/\rho_0$ is sufficiently small to capture accurately all features of the empirical pressures. Consequently, our final results do not change if we make the bin size somewhat larger or smaller.

\begin{figure}[htb]
\begin{center}
  \resizebox{0.5\textwidth}{!}{
  \includegraphics{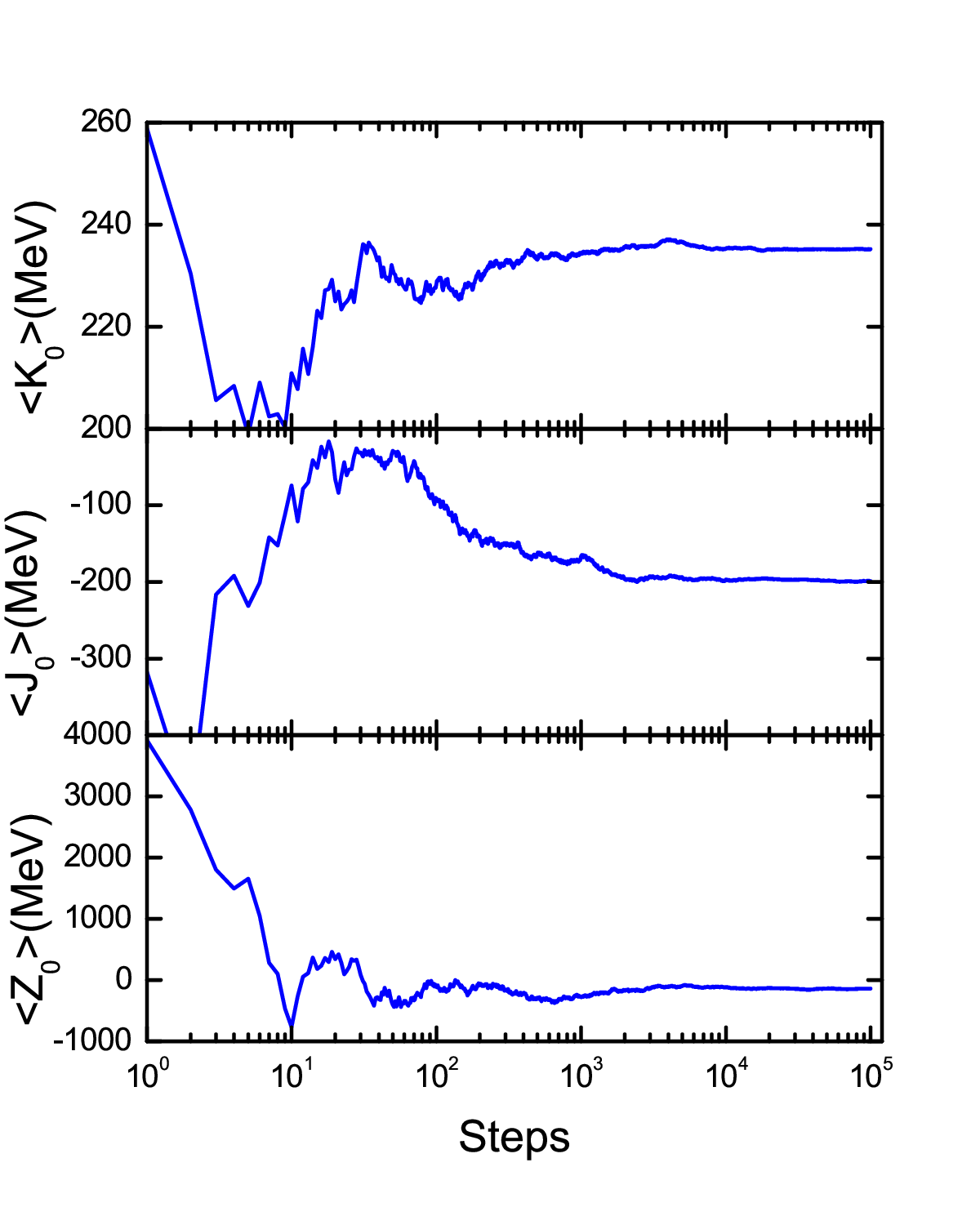}
   }
 \vspace{-0.5cm}
  \caption{(Color online)The mean values of $K_0$ (upper window), $J_0$ (middle window) and $Z_0$ (bottom window) as functions of the step number in the Markov-Chain Monte Carlo sampling of the posterior probability distribution function in the case of setting the likelihood function to 1.}\label{burn-in}
\end{center}
\end{figure}

By using the randomly generated parameters $K_0$, $J_0$ and $Z_0$ as well as the expression (\ref{pressure0}) for SNM pressure, one can construct the model ${\cal M}(K_0,J_0,Z_0)$, i.e. the theoretical value $P_{\mathrm{th},j}$ for the cold SNM pressure. Subsequently, one can calculate the likelihood of this set of parameters according to Eq. (\ref{Likelihood}). The posterior PDF of each parameter is then determined by the marginal estimation, e.g., the PDF for the parameter $K_0$ is given by
\begin{equation}\label{Bay3}
P(K_0|D) = \frac{\int P({\cal M})P(D|{\cal M}) dJ_0dZ_{0}}{\int P(D|{\cal M}) P({\cal M})dK_0dJ_0dZ_{0}}.
\end{equation}

It is well known that some initial samples in the so-called burn-in period may have to be discarded because the MCMC process does not normally sample from the equilibrium (target) distribution in the beginning, see, e.g., Ref. \cite{Trotta17} for more detailed discussions. The length of the burn-in period can be determined by checking the trace plot, i.e., the evolution of the mean values of the parameters as a function of the step number in the MCMC chain. When the chain has reached stationarity, it starts sampling from its equilibrium (target) distribution. Then, both the mean and variance of the trace plot should stay relatively constant \cite{Por}. Shown in Fig. \ref{burn-in} are the mean values of $K_0$ (upper window), $J_0$ (middle window) and $Z_0$ (bottom window) varying with the MCMC steps in the case of setting the likelihood function to 1. It is seen that, after about 20,000 burn-in steps, the means of the three parameters become approximately the means of the prior PDF given in Table \ref{tab:priornew}. In the full calculations with the realistic likelihood function, we discard the 50,000 burn-in steps and use 20 million steps afterwards in calculating the posterior PDFs of $K_0$, $J_0$ and $Z_0$.

\begin{table*}[htbp]
\centering
\caption{\label{tab:mvp}The most probable posterior values and 68\%, 90\% boundaries of $K_0$, $J_0$ and $Z_0$ inferred from using the pressure band constrained by the kaon, flow as well as both kaon \& flow data, respectively. All quantities are in MeV.}
 \begin{tabular}{lccccccc}
  \hline\hline
  Parameters &68\% posterior boundaries &90\% posterior boundaries \\
  &kaon, flow, kaon \& flow &kaon, flow, kaon \& flow \\
  \hline\hline\\
  \vspace{0.2cm}
 $K_0$  &202$^{+14}_{-16}$,  208$^{+16}_{-20}$,  192$^{+12}_{-16}$ &202$^{+24}_{-26}$, 208$^{+28}_{-30}$, 192$^{+22}_{-24}$   \\
 $J_0$  &-260$^{+110}_{-160}$, -240$^{+110}_{-120}$, -180$^{+100}_{-110}$   &-260$^{+200}_{-240}$, -240$^{+180}_{-190}$, -180$^{+170}_{-170}$\\
 $Z_0$  &-200$^{+750}_{-1000}$, 250$^{+300}_{-250}$, 200$^{+250}_{-250}$   &-200$^{+1800}_{-1700}$, 250$^{+500}_{-400}$, 200$^{+400}_{-400}$\\\\
 \hline
 \end{tabular}
\end{table*}

\begin{figure*}[htb]
\begin{center}
\resizebox{0.49\textwidth}{!}{
  \includegraphics{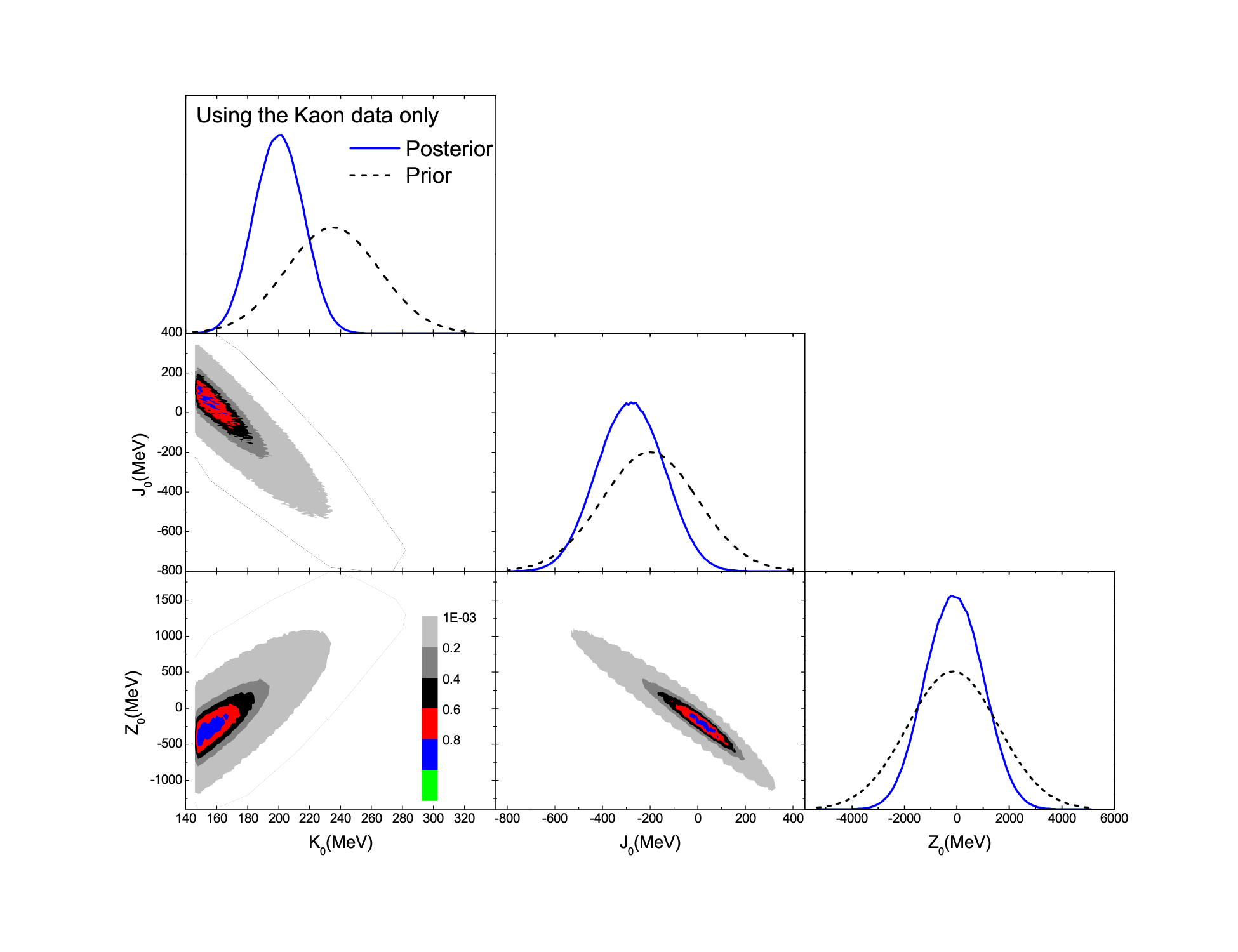}
  }
   \resizebox{0.49\textwidth}{!}{
    \includegraphics{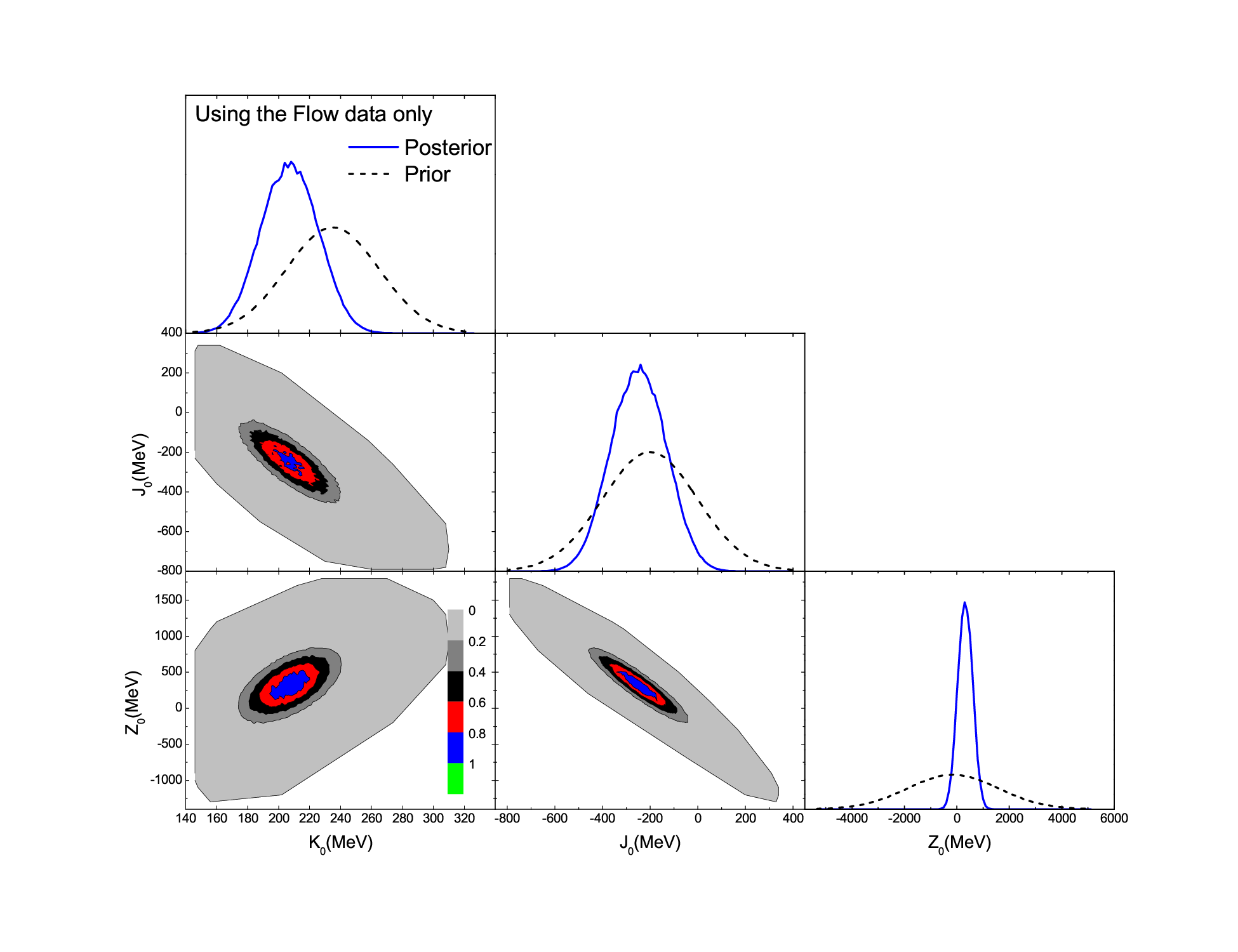}
  }
  \vspace{-0.8cm}
  \resizebox{0.95\textwidth}{!}{
  \includegraphics{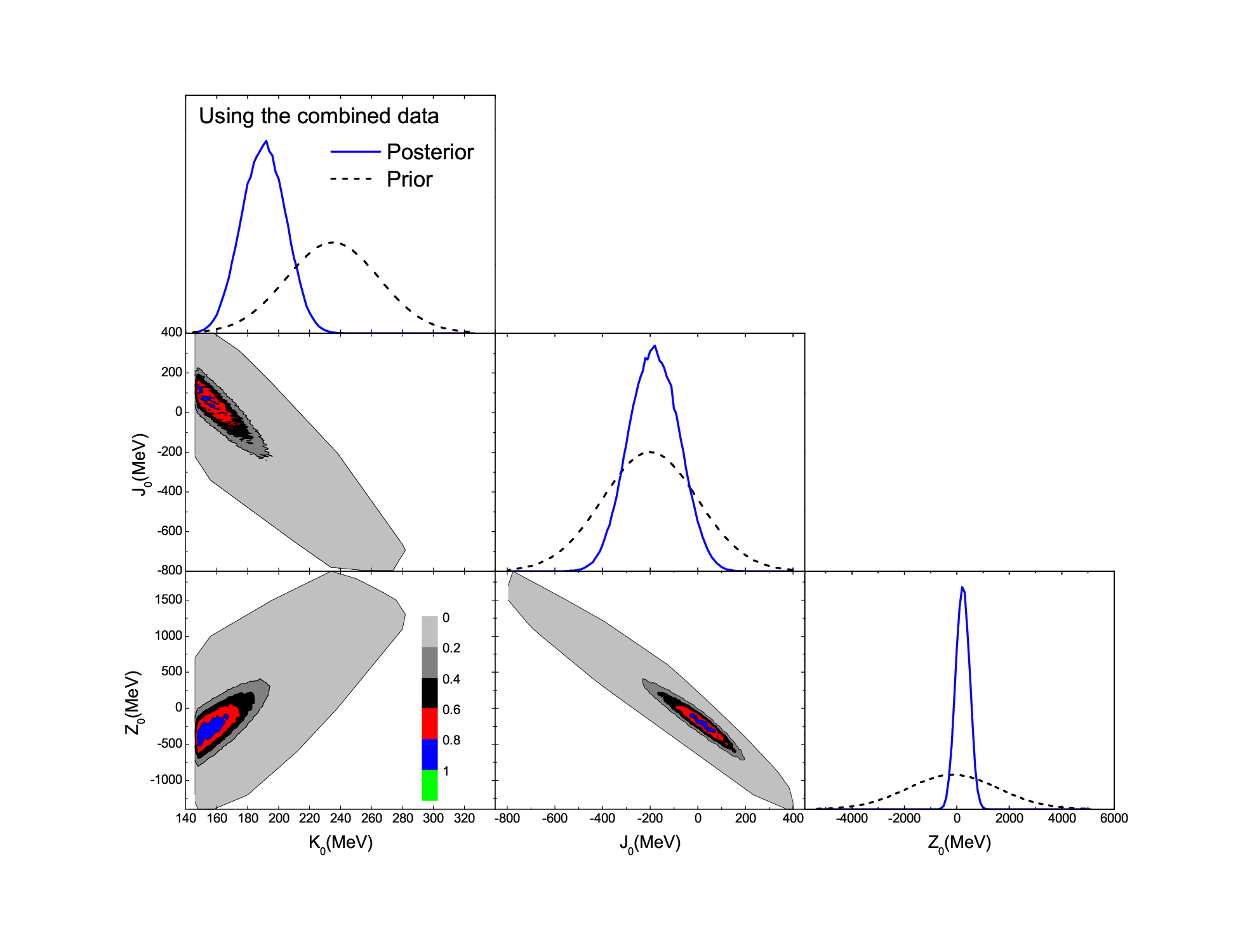}
  }
 \vspace{-0.5cm}
\caption{(Color online)The posterior PDFs for $K_0$, $J_0$ and $Z_0$ and their correlations obtained from the Bayesian analyses of the constraining bands on the SNM pressure shown in Fig. \ref{pressure}. The upper-left, upper-right and lower windows are the results of using the pressure bands from the kaon data, flow data and the combined data, respectively.}\label{kaon-fig3}
\end{center}
\end{figure*}
\section{Results and Discussions}\label{Res}
The 68\% (90\%) credible region for the posterior PDF of each parameter, i.e., the so-called the highest posterior density (HPD) interval \cite{Turkkan14}, is calculated according to
\begin{equation}\label{HPD}
  \int_{p_{i\mathrm{L}}}^{p_{i\mathrm{U}}}\mathrm{PDF}(p_{i})dp_{i}=0.68~ (0.90),
\end{equation}
where $p_{i\mathrm{L}}$ ($p_{i\mathrm{U}}$) is the lower (upper) limit of the corresponding HPD interval of the parameter $p_i$. In the following, we present and discuss results of using the default Gaussian priors and the uniform priors for the three parameters, separately.

\subsection{The posterior PDFs from using the Gaussian priors for all three EOS parameters}
The most probable values of $K_0$, $J_0$ and $Z_0$ together with their 68\% and 90\% credible boundaries  are listed in Table \ref{tab:mvp}. Their posterior and prior PDFs as well as correlations are shown in the upper left (kaon only), upper right (flow only) and lower (combined kaon and flow constraints) windows of Fig. \ref{kaon-fig3}, respectively. Several interesting physics observations can be made:
\begin{itemize}
\item
As listed in Table \ref{tab:mvp}, smaller values of $K_0$ compared to its prior mean are preferred in all cases considered here, while the most probable values of $J_0$ and $Z_0$, especially for $Z_0$, depend strongly on whether the high-density constraint from the flow experiments is used. There is an inverse correlation between $K_0$ and $J_0$, and between $J_0$ and $Z_0$ but a weakly positive correlation between $K_0$ and $Z_0$. These correlations are easily understood from the expression of pressure in Eq. (\ref{pressure0}). Namely, two terms next to each other compensate each other in reproducing the same pressure data under the same condition, they are thus negatively correlated. Consequently, two terms separated by a middle-term (i.e., $J_0$ is between $K_0$ and $Z_0$) are weakly positively correlated. Similar correlations were observed in our earlier studies using neutron star observables \cite{Xie19,Xie20}.

 \item
The constraining band on the SNM pressure in the density range of 1.3$\rho_0$ to 2.2$\rho_0$ alone from the kaon data, as shown in the upper left window of Fig. \ref{kaon-fig3}, constrains significantly the $K_0$ parameter but not the $J_0$ and $Z_0$ parameters relative to their prior PDFs, especially for $Z_0$. More quantitatively, they are only loosely constrained to $J_0=-260^{+110}_{-160}$ MeV and $Z_0=-200^{+1800}_{-1700}$ MeV at 68\% confidence level. This is understandable because the parameter $K_0$, mainly characterizing properties of SNM around its saturation density, plays the dominant role in the density range of 1.3$\rho_0$ to 2.2$\rho_0$. While the parameters $J_0$ and $Z_0$ start affecting significantly the pressure at densities above about 2$\rho_0$ and 3$\rho_0$, respectively. Therefore, the constraining band on the SNM pressure in the density range of 1.3$\rho_0$ to 2.2$\rho_0$ can put a strong limit on $K_0$ but only a weak one on $J_0$ and do not affect much the $Z_0$ parameter.

\item
As shown in the right window of Fig. \ref{kaon-fig3}, the constraining band on the SNM pressure at densities from 2$\rho_0$ to 4.5$\rho_0$ alone from the flow experiments can constrain the $J_0$ and $Z_0$ parameters reasonably tightly to $J_0=-240^{+110}_{-120}$ MeV and $Z_0=250^{+300}_{-250}$ MeV at 68\% confidence level, respectively. They are significantly narrower than those obtained using the kaon experiments only, especially for $Z_0$. However, the posterior PDF of $K_0$ is almost unchanged. This is because in the density range of 2$\rho_0$ to 4.5$\rho_0$ the parameters $J_0$ and $Z_0$ play much stronger roles than the parameter $K_0$ in determining the pressure. Of course, as shown in the lower window of Fig. \ref{kaon-fig3}, combining the constraining bands on the SNM pressure from both kaon production and flow experiments in the whole density range from 1.3$\rho_0$ to 4.5$\rho_0$ leads to even more tighter constrains on all of the parameters, i.e., $K_0=192^{+12}_{-16}$ MeV, $J_0=-180^{+100}_{-110}$ MeV and $Z_0=200^{+250}_{-250}$ MeV at 68\% confidence level, respectively. We notice that the $J_0$ value is consistent but more loosely constrained compared to the value of $J_0=-190^{+40}_{-40}$ MeV from our recent Bayesian analysis \cite{Xie19} of canonical neutron star properties from X-ray and gravitational wave observations using the same prior PDF. Of course, this comparison is somewhat unfair since the $Z_0$ term is not considered in the analysis of neutron star properties in Ref. \cite{Xie19}.

\item The narrowing down of the $Z_0$ parameter by the combined data compared to its broad prior range is rather robust. As shown in the lower window of Fig.\ 3, the posterior PDF of $Z_0$ approaches zero when $Z_0$ is far away from its most probable value $Z_0=200$ MeV. The prior range of $Z_0$ is from -5330 MeV to 5038 MeV whereas the obtained 90\% credible interval of its posterior PDF is between -200 MeV and 600 MeV. This is because in the high density region considered here, both the $J_0$ and $Z_0$ terms contribute to the pressure significantly. The constraining band on the pressure narrows down dramatically the prior PDF of $Z_0$ through the likelihood function, leading to a relatively sharp peak in its posterior PDF. The extracted constraint $Z_0=200^{+250}_{-250}$ MeV at 68\% confidence level, albeit still having a relatively large width, is much more tighter than those from the Skyrme-Hartree-Fock calculations (-903 MeV $\leq Z_0 \leq$ 2128 MeV)\cite{MM1}, the empirical local density functional model (-4478.35 MeV $\leq Z_0 \leq$ -353.91 MeV)\cite{MM1}, the relativistic mean field (2014 MeV $\leq Z_0 \leq$ 9997 MeV) and the relativistic Hartree-Fock (4581 MeV $\leq Z_0 \leq$ 6703 MeV) as mentioned in the introduction \cite{MM1}.

\item
As summarized in Table \ref{tab:mvp}, by combining the kaon and flow data the resulting 68\% or 90\% confidence boundaries of all three EOS parameters are narrowed down compared to the results from analyzing the two data sets independently. Moreover, the observed reduction in uncertainties by combining the two data sets is not simply what one would natively expect from the relationship of $\sqrt{err_1^2+err_2^2}/2$ of two independent measurements with an error bar of $err_1$ and $err_2$, respectively. This is completely understandable as the two data sets cover completely different density ranges as we stated clearly earlier. In fact, as mentioned above, the kaon data in the lower density region constrains strongly the $K_0$ and $J_0$ but weakly the $Z_0$, while it is the other way around for the flow data in the higher density region. Moreover, to our best knowledge, the confidence boundaries in Bayesian statistics should not be expected to follow the rules governing error bars in frequentist statistics.

\end{itemize}

\subsection{Effects of the prior PDFs}
According to the Byes's theorem of Eq. (\ref{Bay1}), the posterior PDFs of the parameters are closely related to their prior distributions. Usually, one can change the prior distributions of the parameters by changing their ranges and/or shapes. In the following, we discuss effects of using different prior PDFs on the posterior PDFs. For comparisons with the default results we switch the Gaussian PDF to a uniform one for one parameter each time while keep the default prior PDFs for the remaining two parameters.

\begin{figure*}[htb]
\begin{center}
\resizebox{0.8\textwidth}{!}{
  \includegraphics{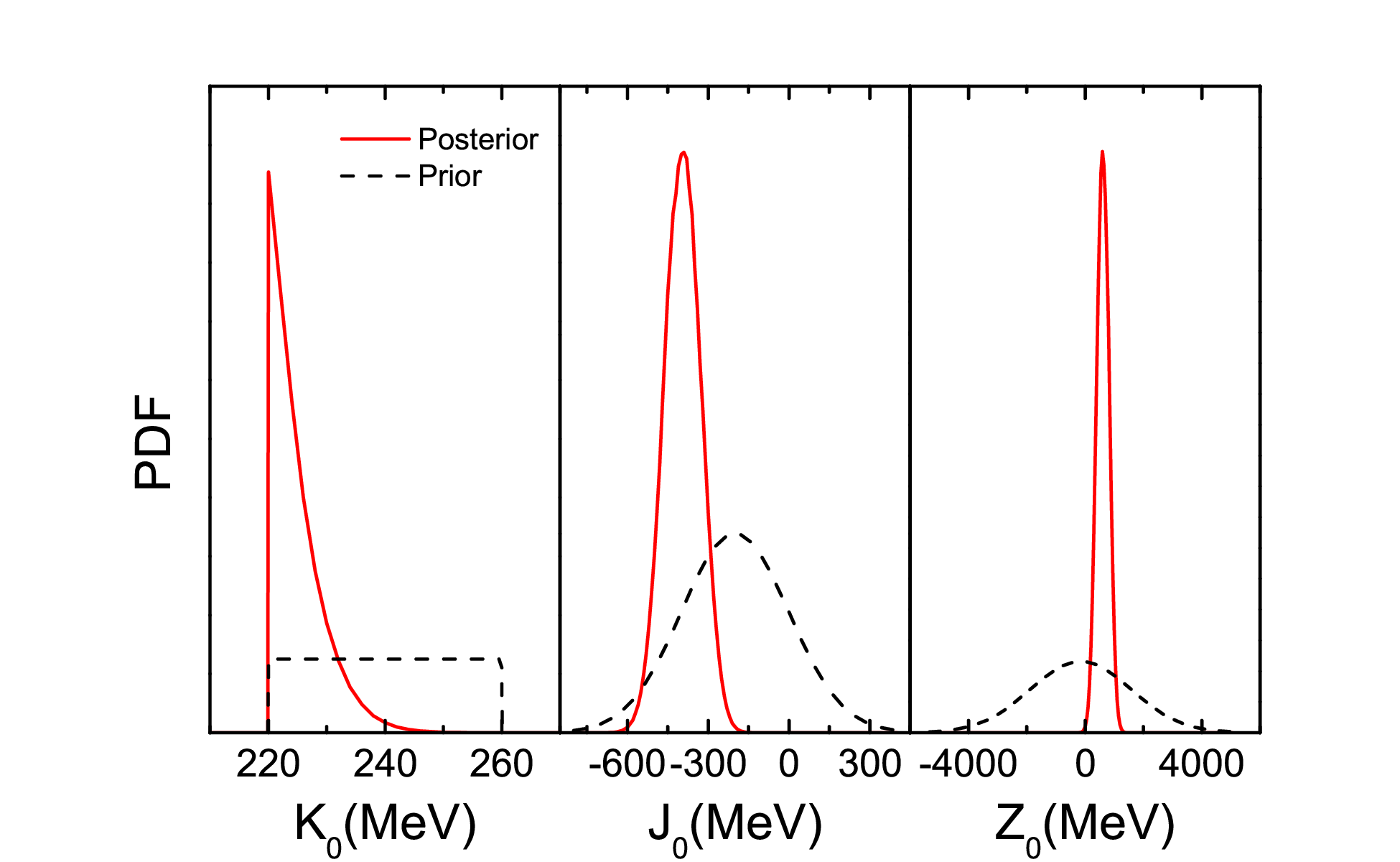}
  }
\caption{(Color online) The posterior PDFs of $K_0$, $J_0$ and $Z_0$ from the Bayesian analyses of the constraining bands on the SNM pressure from the combined data. The default Gaussian prior PDFs for $J_0$ and $Z_0$ while a uniform prior PDF for $K_0$ between 220 MeV and 260 MeV are used as indicated by the dashed lines.}\label{K0fig7}
\end{center}
\end{figure*}

First, lets examine effects of  the prior PDF of $K_0$. The value of $K_0$ is much better known than the other two parameters. As the default, we have used above the Gaussian function with the mean and $1\sigma$ variance of $235\pm 30$ as the prior PDF for $K_0$. Given the long history and diverse results of extracting the $K_0$ from various experiments using different approaches, this choice is physically sound \cite{MM1}. 
However, one can easily find in the literature many evidences/conclusions, see, e.g., Refs. \cite{shlomo06,Garg18,Colo14}, indicating/stating clearly that the $K_0$ has an absolute range of 220 MeV to 260 MeV, see, e.g., the statement ``The conclusion that the $K_0$ should be in the range of $240\pm 20$ MeV has been reached since about one decade \cite{shlomo06}" on page 80 of Ref. \cite{Garg18} and 
Fig. 8 by some of the same authors in Ref. \cite{Colo14} where this range was used as the absolute boundary of $K_0$ in excluding many model predictions. Since we can not find any other explicit statement regarding the confidence level of the error bars in $K_0$ in Refs. \cite{shlomo06,Garg18,Colo14}, we regard the choice of using a uniform prior between 220 MeV and 260 MeV as a strong belief with 100\% confidence. 
In fact, such a strong belief has been widely used in the literature to judge model predictions and in astrophysical applications. In our opinion, it would be simply ignorant to disregard this choice as part of our current knowledge about $K_0$. While we do not have a preference for one of the two choices discussed here, it is interesting to know how the strong belief about the range of $K_0$ may affect what we extract from the Bayesian analysis of the nuclear pressure from heavy-ion collisions. For this purpose, we compare the default results with a calculation assuming $K_0$ has a uniform prior in the range of 220 to 260 MeV while keeping the Gaussian priors for $J_0$ and $Z_0$ in their original ranges. Even without any new calculation, based on the Baye's theorem and the default results shown in Fig. \ref{kaon-fig3}, one expects this choice will lead to a sharp cut-off of the posterior PDF of $K_0$ at 220 MeV, and it may also have some secondary effects on the PDFs of $J_0$ and $Z_0$ because of their correlations with $K_0$.

Shown in Fig. \ref{K0fig7} are the posterior PDFs for $K_0$, $J_0$ and $Z_0$ from the combined data by using a uniform prior for $K_0$ between 220 MeV and 260 MeV but the default Gaussian priors for $J_0$ and $Z_0$. Compared to the default results shown in the lower window of Fig. \ref{Res}, the change in the posterior PDF of $K_0$ is what we expected. It now peaks sharply at 220 MeV instead of the Gaussian shaped posterior PDF peaked at 192 MeV in the default calculation. Obviously, this is a strong effect of the strong belief in the range of $K_0$. It is also seen that the PDF of $J_0$ shifts to more negative values while that of $Z_0$ shift to higher values. More quantitatively, at 68\% credible level, the most probable values of $K_0$, $J_0$ and $Z_0$ are now $K_0=220^{+6}_{-0}$ MeV, $J_0=-390^{+60}_{-70}$ MeV and $Z_0=600^{+200}_{-200}$ MeV, respectively.  Compared to their default values listed in Table \ref{tab:mvp}, their credible ranges become tighter besides the obvious shifts in their most probable values. These results clearly demonstrate the importance of the prior PDFs. They can all be easily understood: (1) By limiting the $K_0$ to 220 to 260 MeV, the influence of the pressure bands near the saturation density from kaon production is significantly reduced. As discussed in Refs. \cite{Fuchs,lynch09,kaos1,kaos2},
transport model analyses of kaon production experiments favor a soft EOS with $K_0=200$ MeV. Setting the prior range of $K_0$ between 220 and 260 MeV essentially cuts off the main influence of the kaon data in sampling its posterior PDF. Effectively, the average value of $K_0$ is increased compared to its default value. Since the $K_0$ is negatively correlated to $J_0$ but positively correlated to $Z_0$ as we discussed earlier, the posterior PDF of $J_0$ is shifted to more negative values while that of $Z_0$ is shifted to more positive values to conserve the total pressure. (2) The more authoritarian prior in a smaller absolute range between 220 and 260 MeV for $K_0$ but giving every value a more democratic treatment naturally leads to the more tighter posterior bounds on all three EOS parameters compared to their default values. 
\begin{figure*}[htb]
\begin{center}
\resizebox{0.8\textwidth}{!}{
  \includegraphics{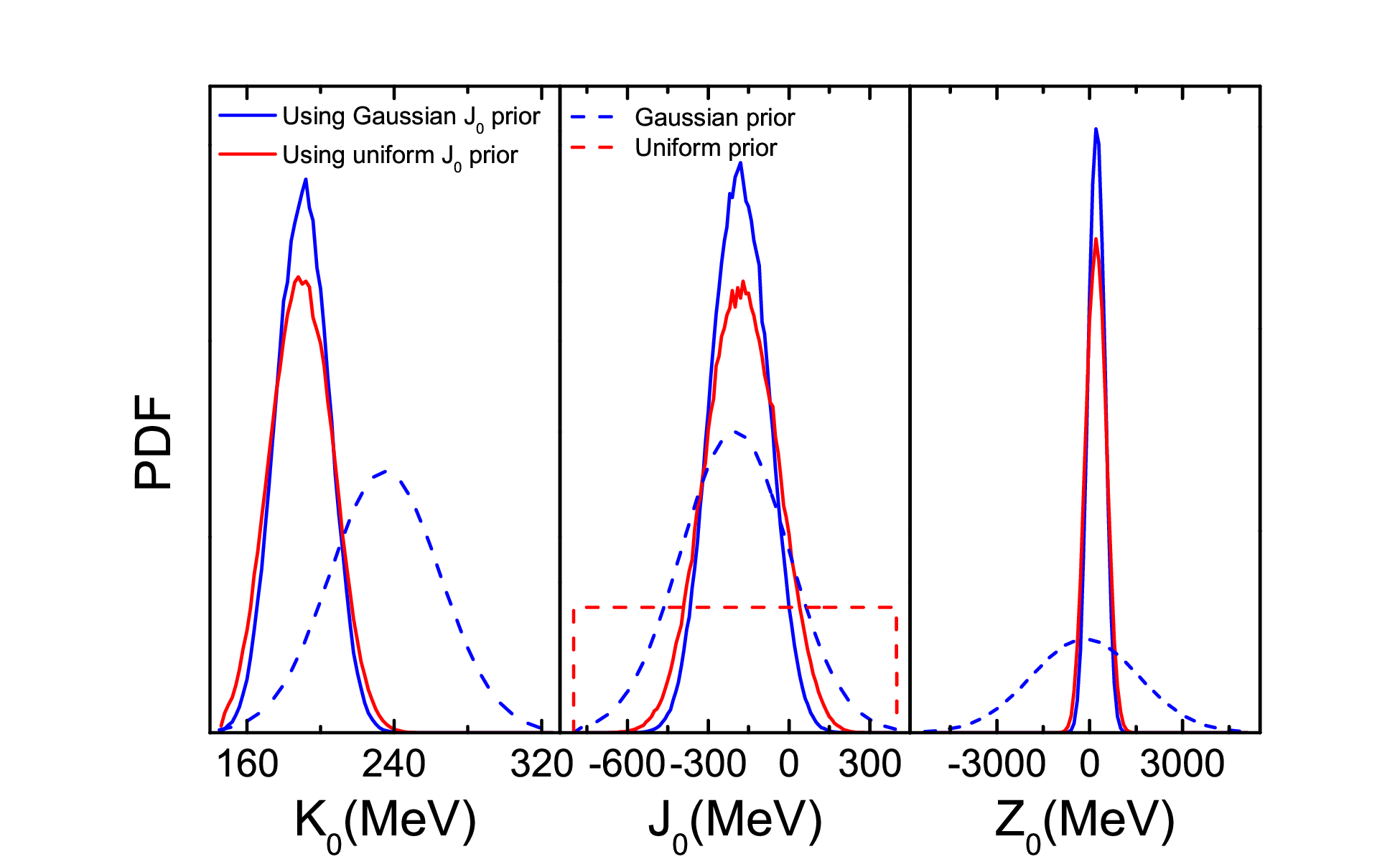}
  }
  \resizebox{0.8\textwidth}{!}{
  \includegraphics{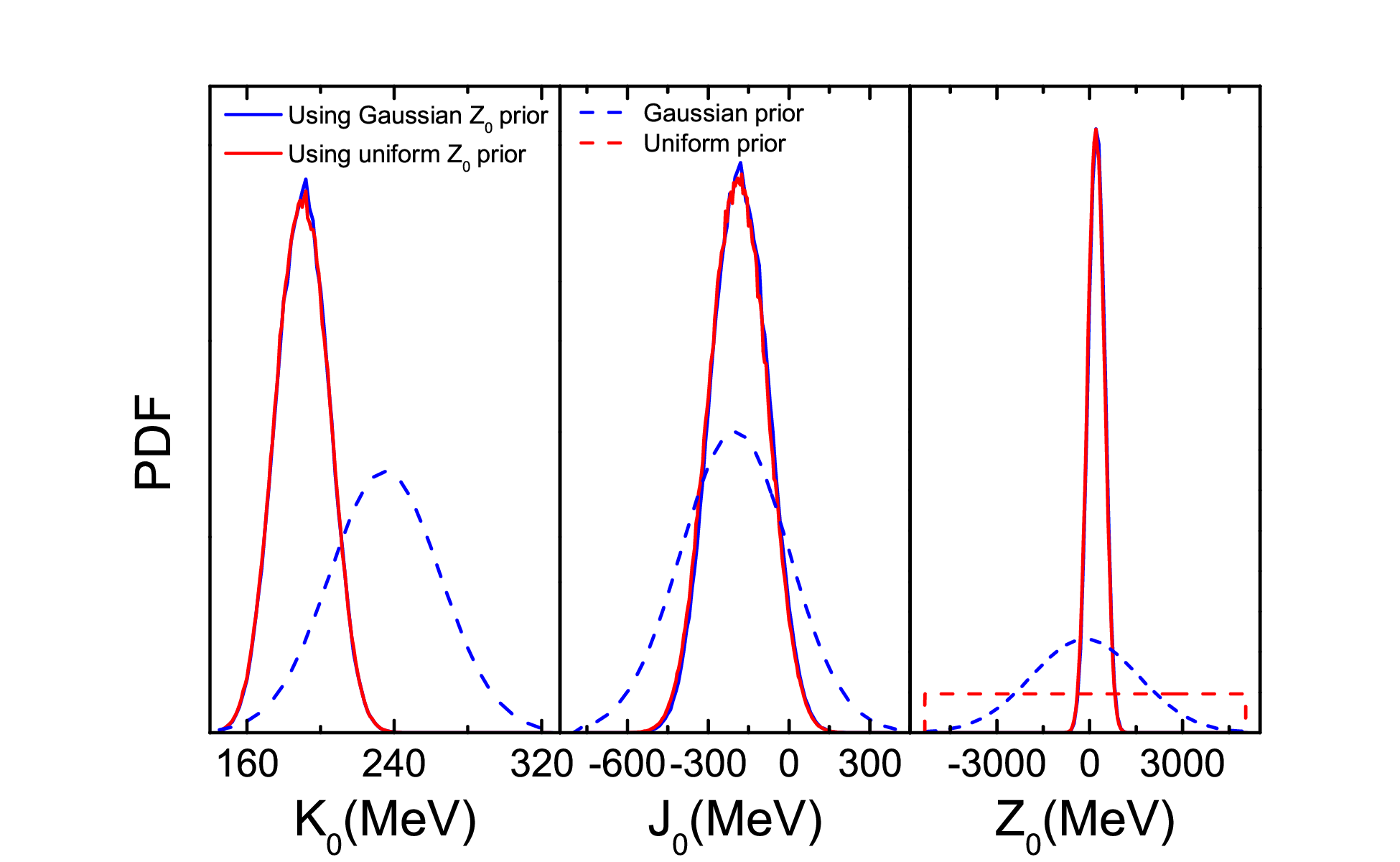}
  }
\caption{(Color online) (Upper) The posterior PDFs for $K_0$, $J_0$ and $Z_0$ using Gaussian prior PDFs for $K_0$ and $Z_0$ but both uniform (red) and Gaussian (blue) priors for $J_0$. (Lower): Results using Gaussian prior PDFs for $K_0$ and $J_0$ but both uniform (red) and Gaussian (blue) priors for $Z_0$. }\label{J0effect-fig6}
\end{center}
\end{figure*}

Unlike the situation for $K_0$, to our best knowledge, for both $J_0$ and $Z_0$ there is not really any reliable information about the shapes of their prior PDFs besides their rough ranges. To see effects of their prior PDFs on the posterior PDFs, we have done calculations using uniform priors for $J_0$ and $Z_0$ in the same default ranges listed in Table \ref{tab:priornew}.
Shown in Fig. \ref{J0effect-fig6} are comparisons of the posterior PDFs for $K_0$, $J_0$ and $Z_0$ using the default priors and the uniform priors for $J_0$ and $Z_0$, respectively.
It is seen that the variations of the prior PDFs for $J_0$ and $Z_0$ have little influence on the posterior PDFs of $K_0$, $J_0$ and $Z_0$, especially for the variation of the $Z_0$ prior. This is because in the density region of 1.3$\rho_0$ to 4.5$\rho_0$ the constraining band on pressure from heavy-ion collisions through the likelihood function is so strong that the changes in the prior PDFs of $J_0$ and $Z_0$ do not lead to significant modifications to the posterior PDFs of any of the three EOS parameters.

In short, among the three EOS parameters considered, only the variation of the prior PDF of $K_0$ has a large effect on the posterior PDFs of the three EOS parameters. This is mainly because the pressure in the density range of 1.3$\rho_0$ to 4.5$\rho_0$ do not constrain the $K_0$ (characterizing the EOS around $\rho_0$) as strongly as the $J_0$ and $Z_0$. Nevertheless, the posterior PDFs from the analyses using different priors especially for $J_0$ and $Z_0$ are qualitatively consistent. Moreover, their posterior uncertainties are also much smaller than their current values, representing a significant improvement to our current knowledge.

\section{Summary and outlook}
In summary, adopting the constraining bands on the pressure in cold SNM in the density range from 1.3$\rho_0$ to 4.5$\rho_0$ from analyzing relativistic heavy-ion collisions we inferred the posterior PDFs of the underlying SNM
incompressibility $K_0$, skewness $J_0$ and kurtosis $Z_0$ parameters within the Bayesian framework using a parameterized EOS.  Assuming the three parameters have Gaussian priors centered around $235\pm 30$, $-200\pm 200$ and $-146\pm 1728$ MeV at 68\% confidence level, their posterior most probable values are found to be $K_0$=192$^{+12}_{-16}$ MeV, $J_0$=-180$^{+100}_{-110}$ MeV and $Z_0$=200$^{+250}_{-250}$ MeV, respectively. We also found that the pressure band  between 1.3$\rho_0$ to 4.5$\rho_0$ is so constraining on $J_0$ and $Z_0$ that the variations of their prior PDFs do not affect much the posterior PDFs of the three EOS parameters.
However, the variation of the prior PDFs of $K_0$ has significant effects on the posterior PDFs. In particular, adopting the strong belief that $K_0$ has an equal probability within its absolute boundary of 220 MeV to 260 MeV widely used in the literature, the posterior most probable values of the three parameters shift to $K_0=220^{+6}_{-0}$ MeV, $J_0=-390^{+60}_{-70}$ MeV and $Z_0=600^{+200}_{-200}$ MeV, respectively. Despite of the
dependence on the prior PDF of $K_0$, the resulting posterior PDFs are all consistent and understandable. Moreover, their posterior uncertainties are significantly smaller than their current values especially for the skewness $J_0$ and kurtosis $Z_0$ parameters.

The resulting most probable value of $K_0$ from heavy-ion collisions is slightly lower than that from analyzing giant resonances while their uncertainty ranges overlap. This is not surprising as the extraction of cold EOS parameters from observables of heavy-ion reactions depends on how accurately we known about the in-medium elementary nuclear reaction cross sections as we discussed earlier. While the extraction of $K_0$ from giant resonances depends on the correlation of $K_0$ with the poorly known density dependence of nuclear symmetry energy. Nevertheless, as we demonstrated in this work, combining the knowledge from both areas allowed us to narrow down significantly the skewness $J_0$ and kurtosis $Z_0$ parameters of high-density SNM EOS.

There is an obvious ambiguity about what prior PDF for $K_0$ one should use given its
quantitative effects on the posterior PDFs of the EOS parameters in our analysis.
While one may never completely get away from strong beliefs and possible biases
because of our limited knowledge, perhaps one way to reduce the ambiguity is to
perform a comprehensive Bayesian analysis of all the original giant resonance data
using several approaches available in the literature. To our best knowledge, such an
analysis has not been done yet. On the other hand, our Bayesian analysis here used
the constraining band of the pressure in cold SNM extracted from transport model
analyses of the kaon production and nuclear collective flow in relativistic heavy-ion
collisions. In doing so, we used the pressure bands as empirical data instead of the
original data in heavy-ion collisions. Ideally, one would use the transport models within
the Bayesian statistical framework and then extract directly from the actual data of
heavy-ion collisions the posterior PDFs of all model parameters including the EOS
parameters, in-medium nuclear cross sections and other transport model parameters. Such an approach has been used successfully in extracting
QGP properties from ultra-relativistic heavy-ion collisions at RHIC and LHC energies,
see, e.g., Ref. \cite{Bernhard19}. Our efforts in this direction are ongoing.

\section*{Acknowledgement} We would like to thank Bao-Jun Cai and Lie-Wen Chen for helpful discussions. WJX acknowledges the National Natural Science Foundation of  China under Grant No. 11505150.
BALI is supported in part by the U.S. Department of Energy, Office of Science, under Award Number DE-SC0013702 and the CUSTIPEN (China-U.S. Theory Institute for Physics with Exotic Nuclei) under the US Department of Energy Grant No. DE-SC0009971.

\clearpage
\end{document}